\documentclass[manuscript]{aastex}

\shorttitle{Distinct states of variability}
\shortauthors{Meyer et al.}

\begin{document}

\title{A formal method for identifying distinct states of variability in time-varying sources: Sgr~A* as an example}

\author{L. Meyer, G.Witzel}
\affil{Department of Physics and Astronomy, University of California,
    Los Angeles, CA 90095-1547}
\author{F. A. Longstaff}
\affil{UCLA Anderson School of Management, University of California,
    Los Angeles, CA 90095-1481}
\author{A. M. Ghez}
\affil{Department of Physics and Astronomy, University of California,
    Los Angeles, CA 90095-1547}

\begin{abstract}
Continuously time variable sources are often characterized by their power spectral density and flux distribution. These quantities can undergo dramatic changes over time if the underlying physical processes change. However, some changes can be subtle and not distinguishable using standard statistical approaches. Here, we report a methodology that aims to identify distinct but  similar states of time variability. We apply this method to the Galactic supermassive black hole, where 2.2 $\mu$m flux is observed from a source associated with Sgr~A*, and where two distinct states have recently been suggested. Our approach is taken from mathematical finance and works with conditional flux density distributions that depend on the previous flux value. The discrete, unobserved (hidden) state variable is modeled as a stochastic process and the transition probabilities are inferred from the flux density time series. Using the most comprehensive data set to date, in which all Keck and a majority of the publicly available VLT data have been merged, we show that Sgr~A* is sufficiently described by a single intrinsic state. However the observed flux densities exhibit two states: a noise-dominated and a source-dominated one. Our methodology reported here will prove extremely useful to assess the effects of the putative gas cloud G2 that is on its way toward the black hole and might create a new state of variability.

\end{abstract}


\section{Introduction}

Many different mechanisms can cause an astronomical source to be variable. Accreting black holes, for example, are variable electromagnetic sources for reasons such as oscillations in an accretion disk, changes in the accretion rate, or turbulent plasma processes like magnetic reconnection leading to an acceleration of electrons. Most often the observed variability in the flux can be well described by a single stochastic process such as a random walk \citep[e.g.][]{kelly09, macleod10, zu13}. If, however, the accretion rate suddenly jumps (e.g. due to the tidal disruption of a star or asteroid) the observed variability can change drastically and in a way that is not well described by a random walk. Such a state change in the variability could be obvious if the effects are large like an increase of the mean flux by orders-of-magnitude. However, there could be changes that are subtle and not trivial to detect. In this paper, we present a formal method to do just that. We apply the method to the massive black hole in the center of the Milky Way. This source is of particular interest, since two distinct states have been claimed to be present in the past \citep{dodds11}, and the upcoming encounter with the gaseous object G2 \citep{gillessen12,gillessen13a,gillessen13b,phifer13,meyer13} might or might not lead to a variability state change.      

The emission associated with the accretion flow around the Galactic black hole, Sgr~A*, has been detected in a few wavelength regimes \citep[for recent reviews see][]{genzel10,morris12,falcke13}. While it was first detected at radio wavelengths a few decades ago \citep{balick74}, its discovery in the X-rays \citep{baganoff01} and near-infrared \citep{genzel03, ghez04,eckart04} did not happen until the early 2000s when advanced imaging systems came online (Chandra/XMM in the X-rays and adaptive optics in the near-infrared).  Sgr~A* is in Eddington terms the most under-luminous massive black hole accessible to observations. Its unexpected faintness inspired a class of radiative inefficient accretion flow models \citep[e.g.][]{narayan95,blandford99,yuan1,yuan2}.  

While Sgr~A* is variable across all observable wavelengths, its variability is most pronounced in the  near-infrared (NIR) and X-rays with flux excursions that lie a factor of $\sim10$ (for the NIR) and $\sim100$ (for the X-ray), respectively, above the low flux levels. Early studies of Sgr~A*'s NIR and X-ray variability reported the existence of a quasi-periodic oscillation (QPO) of $\sim17$ min in the flux \citep{genzel03,aschenbach04}. However, this finding was not confirmed by later statistical analyses \citep{do09}. The potential existence of a QPO was met with great interest, since it potentially offers a way to measure the spin of the black hole and to test the curvature of space-time close to it \citep[e.g.][]{broderick05,meyer06b,johannsen11}.

One advantage of the NIR over the X-ray regime for studying the variability of Sgr~A* is that it is visible at NIR wavelengths much more of the time. While Sgr~A*'s X-ray emission peaks above the steady background created by the extended, larger-scale thermal accretion flow around $4 \%$ of the time \citep{neilsen13}, Sgr~A*'s NIR emission is almost always ($\gtrsim 90 \%$) detected at the highest angular resolution possible today with Keck Observatory (as we will show here and in Witzel et al., in prep.). Recent NIR studies have shown that  Sgr~A* is sufficiently modeled as a purely random process.   The power-spectral density (PSD) of Sgr~A*'s time variability is a featureless power-law for relatively high frequencies, a characteristic that can be modeled with a random walk and is more generally called red noise \citep{do09,meyer08}. At lower frequencies, the power-law breaks to a shallower slope at a timescale of $150 - 600$\,minutes \citep{meyer09,witzel12}.

In addition to the PSD, another key quantity to describe Sgr~A*'s variability is its flux density distribution. Recently, \cite{dodds11} and \citet{witzel12} looked at the total flux density distribution in the NIR by constructing a histogram of all observed flux density values. Interestingly, their interpretations are quite different: \cite{dodds11} use a log-normal distribution + a power-law tail -- convolved with a Gaussian to account for measurement errors -- to describe the distribution and argue that Sgr~A* has two distinct states, one described by the log-normal part, the other by the power-law tail. In contrast, \cite{witzel12} find that only a power-law (convolved with a Gaussian) is needed to accurately describe the flux density histogram of Sgr~A*. It is important to note that both studies do not use timing information to argue for one or two states. The question whether or not multiple states can be inferred from the data was identified as a key question in understanding Sgr~A* by \citet{genzel10}.  

In this work, we derive a formal method to answer the question of how many states can be inferred from Sgr~A*'s light curve.  In section 2 of the paper we will present our methodology in detail. While it has been developed with the specific case of Sgr~A* in mind, its application should be more general. It is useful whenever one wants to investigate if a time variable source exhibits distinct states of variability. This method is known in economics as ``regime switching model'' \citep[see, e.g.,][]{hamilton}. We then apply this approach to Sgr~A* using the most extensive NIR data set constructed to date (Witzel et al., in prep.). We end by discussing how to assess the upcoming impact of the gaseous object G2 on the accretion flow around the black hole. Our methodology along with the unprecedented data set, which represents the best base line of Sgr~A*'s behavior before G2 swings around in early 2014, is the ideal tool set to quantify Sgr~A*'s response to any mass accreted from G2.

\section{Methodology}

\begin{figure}
\includegraphics[scale=.75]{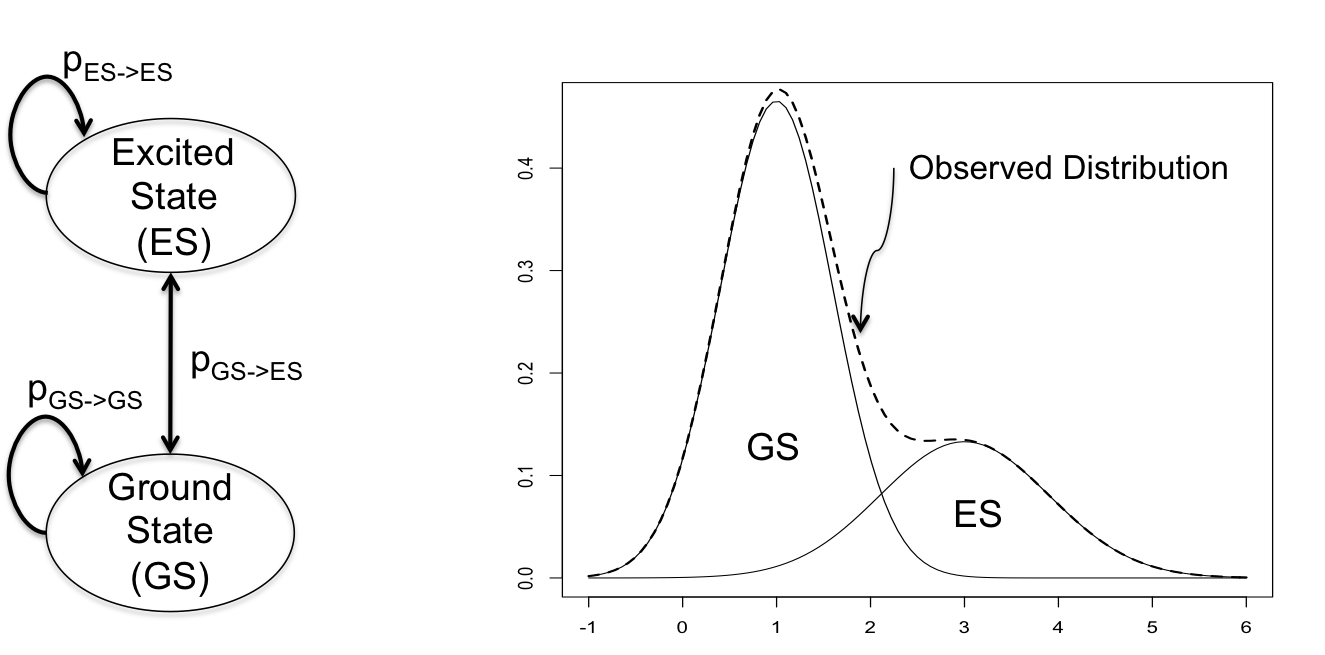}
\caption{A sketch of a 2-state hidden Markov model. The observable quantity shows the dashed overall distribution. It can be decomposed into a ground state (GS) and an excited state (ES), both of which are not directly observable. Given a sequence of observations, the parameters of the individual distributions as well as the transition probabilities can be solved for. Note that this sketch is overly simplified and in our application to time series the overall, unconditional distribution can not simply be calculated as the weighted sum of the conditional ones. }
\label{doubleGauss}
\end{figure}

The notion of hidden states in a time series has been applied to many diverse fields for a few decades \citep[e.g][]{rabiner89}. Simply put, the idea is that observables entail information about states  that are not directly accessible to the observer. Fig.~\ref{doubleGauss} sketches a simple system that can be in a ground state and an excited state and the distribution of the observable is different for each state. Overall, a mixture distribution would be observed. A 2-state hidden Markov model that looks at the time sequence of observations and assumes a general form for the distributions is able to determine their parameters, the probability to transition from the ground to the excited state and vice versa, as well as the probability to be in a certain state for every observation. Modeling the stochastic process of the state variable as Markovian means that the process satisfies the Markov property: one can make predictions for the future state based solely on its present state just as well as one could knowing the process' full history. 

The key to this approach is that we move beyond the simple unconditional flux distribution -- measured by the histogram of all flux densities -- and use the information in the time series to identify conditional flux distributions. To illustrate the difference between unconditional and conditional distributions, consider the following analogy. Imagine that one wanted to forecast the high temperature on the Santa Monica beach tomorrow. One way to do this would be to compute the distribution of all high temperatures over the entire year and use the mean of this unconditional distribution as the forecast. A much better way to proceed, however, would be to estimate the distribution of day-ahead temperatures conditional on today's high temperature being, say, 50 degrees. By conditioning on today's temperature, the distribution for tomorrow's temperature would be much tighter and more informative that would be the unconditional distribution based on the entire year. This analogy illustrates the intuition behind our approach. We use the current flux measurement to specify the conditional distribution of the subsequent flux measurement, and estimate a range of conditional distributions, one for each state, rather than just using the single unconditional distribution.

In the following, we will first describe our approach and the construction of the likelihood function in more detail with a simplified  example \citep[adopted from][]{hamilton05} and then generalize this to the more complex case of Sgr~A*.

\subsection{Simple example}

Let us consider a hypothetical flux time series where $y_t$ is the flux measured at time $t$, and the data are well described by a stochastic model of the form
\begin{equation}
y_t = c_1 + \phi y_{t-1} + \epsilon_t,
\end{equation}
where $c_1$ and $| \phi | < 1$ are constants, and $\epsilon \sim \mathcal{N} (0,\sigma^2)$. Let us further assume that we would like to test whether a model that allows a change in mean flux  gives a better description of the data. For a permanent change of the mean flux we could just write down a different parameter $c_2$ for some $t > t_1$,
\begin{equation}
y_t = c_2 + \phi y_{t-1} + \epsilon_t.
\end{equation}
We are, however, interested in a situation where we can jump back and forth between both models. We will refer to the different models as different states, and at a given time the source emitting the flux is in state $s_t  = j$, where $j = 1$ or $2$ in our example. We will assume that this state is not directly observable (e.g. with spectroscopic observations), but has to be inferred from the light curve itself. 

The unobserved state is modeled using a Markov switching approach. Concretely, let $p_{ij}$ be a matrix that reflects the transition probabilities of switching to another state or remaining in the current state,
\begin{equation}
p_{ij} = \textnormal{Prob}(s_t = j | s_{t-1} = i),
\end{equation}
and let $\Pi_{jt}$ be the probability of being in state $j$ at time $t$,
\begin{equation}
\Pi_{jt} = \textnormal{Prob}(s_t = j | y_t).
\end{equation}
The probability density of $y_t$ for our model is then
\begin{equation}
f(y_t | s_t=j,y_{t-1}) = \frac{1}{\sqrt{2\pi \sigma^2}} \exp\left({\frac{-(y_t-c_j-\phi y_{t-1})^2}{2 \sigma^2}}\right) = f_{jt}.
\end{equation}
With these notations, the conditional probability density of $y_t$ can be written as
\begin{equation}
h(y_t | y_{t-1};\theta) = \sum_{i,j} p_{ij} \Pi_{i t-1} f_{jt},
\end{equation}
where $\theta$ means the set of parameters in the model (here, $c_j$, $\phi$, and $p_{ij}$). Since $\Pi_{jt}$ is not directly observable, it has to be recursively calculated by observing that
\begin{equation}
\Pi_{jt} = \frac{\sum_i p_{ij} \Pi_{it-1} f_{jt}}{h(y_t | y_{t-1}; \theta)}.
\end{equation}
Starting at $t_0$ we can execute the iteration to solve for all $\Pi_{jt}$. This means that the final log-likelihood function of the whole light curve is given by
\begin{equation} \label{likeli}
\log h(y_1,y_2,y_3,...,y_T | y_0; \theta) = \sum_{t=1}^T \log h(y_t | y_{t-1};\theta).
\end{equation}
The maximum of this log-likelihood function gives the preferred values for $c_j$, $\phi$, and $p_{ij}$.

\subsection{Extension to Sgr~A*}

While the simple model elucidates the construction of the likelihood function, it is not well suited to be applied to light curves from Sgr~A*. The main reason is that  the real data set shows unevenly sampled measurements with big gaps representing limited telescope time, the night/day cycle, the observability of the Galactic center from the ground throughout a year, measurements of sky background, and instrument failures. This requires us to use a more appropriate model that is time continuous and not discrete as in the above example. Furthermore, the flux density distribution is not Gaussian, and the data contain measurement noise. This is where the astronomical application of the regime switching approach departs from mathematical finance.  We will deal with these points step-by-step in the following. 

A popular model to describe the random variability of Quasars is a so-called damped random walk, which is the only  process that is stationary, Gaussian, and Markov. This process is also known as an Ornstein-Uhlenbeck (OU) model, which is the nomenclature we will use here. It is similar to red-noise models with a broken power-law as the power spectral density, which have been used to model Sgr~A* in the past \citep{do09,meyer09,witzel12}.  Most recently, \citet{dexter13} used an OU process to successfully describe the sub-mm variability of Sgr~A*. Additionally, \citet{kelly09}, \citet{macleod10}, and \citet{zu13} have shown that an OU process is an excellent description of AGN flux variations. Its key advantage for our purpose is that it is a time continuous model which makes the handling of sampling gaps easier. 

The OU model is determined by three parameters, the mean $\mu$, the speed of mean reversion $k$, and the volatility $\sigma$, and the dynamics can be described by the following stochastic differential equation:
\begin{equation} \label{OU}
dy = k(\mu - y) dt + \sigma dZ_t,
\end{equation}
where $dZ_t$ is the increment of a Brownian motion with $Z_t \sim \mathcal{N} (0,t)$. A solution to this equation is the conditional distribution
\begin{eqnarray}
f(y_{t+\Delta t} | y_t) & = & \frac{1}{\sqrt {2\pi \tilde{\sigma}^2}} \exp\left(\frac{-(y_{t+\Delta t} - \tilde{\mu})^2}{2\tilde{\sigma}^2}\right), \\ [5pt]
\tilde{\mu} & = & y_t e^{-k\Delta t} + \mu \left( 1- e^{-k\Delta t}\right), \\ [5pt]
 \tilde{\sigma}^2 & = & \frac{\sigma^2 (1- e^{-2k\Delta t})}{2k}.
\end{eqnarray}

Our goal is to identify whether more than one state is present in the light curves from Sgr~A*. In order to test a model with two or more states against the baseline model of one state only, it is important that the baseline model is already a good description of the overall flux distribution and power spectral density. A OU process matches the observed characteristics of Sgr~A*'s power spectrum: a broken power-law which is otherwise featureless. The observed flux distribution, however, is not Gaussian. As reported by \citet{witzel12} and \citet{dodds11} a heavy tailed, power-law like distribution convolved with a Gaussian describes the flux density distribution more accurately.  We will therefore also use the exponential and double-exponential of a OU process to model Sgr~A*. Since this translates into the equivalent of  a log-normal (and loglog-normal) distribution, we will in the following use the notation of a  $\log$- and $\log\log$-OU process. Please note that taking the logarithm twice is purely empirically motivated and not rooted in physical considerations. Ideally, we would like to use a power-law distribution, but a stochastic model similar to eq.~(\ref{OU}) which results in a power-law density is not known.  

The conditional distributions for the $\log$- and $\log\log$-OU processes are given as follows: (1) for the $\log$ of a OU process,
\begin{eqnarray} \label{logOU}
f(y_{t+\Delta t} | y_t) & = & \frac{1}{\sqrt {2\pi \tilde{\sigma}^2}y_{t+\Delta t}} \exp\left(\frac{-(\ln{(y_{t+\Delta t})} - \tilde{\mu})^2}{2\tilde{\sigma}^2}\right), \\ [5pt]
\tilde{\mu} & = & \ln{(y_t)} e^{-k\Delta t} + \mu \left( 1- e^{-k\Delta t}\right), \\ [5pt]
 \tilde{\sigma}^2 & = & \frac{\sigma^2 (1- e^{-2k\Delta t})}{2k},
\end{eqnarray}
and (2) for the $\log\log$ of a OU process,
\begin{eqnarray} \label{loglogOU}
f(y_{t+\Delta t} | y_t) & = & \frac{1}{\sqrt {2\pi \tilde{\sigma}^2}y_{t+\Delta t}\ln(y_{t+\Delta t})} \exp\left(\frac{-(\ln(\ln(y_{t+\Delta t})) - \tilde{\mu})^2}{2\tilde{\sigma}^2}\right), \\ [6pt]
\tilde{\mu} & = & \ln(\ln(y_t)) e^{-k\Delta t} + \mu \left( 1- e^{-k\Delta t}\right), \\ [6pt]
 \tilde{\sigma}^2 & = & \frac{\sigma^2 (1- e^{-2k\Delta t})}{2k}.
\end{eqnarray}

The choice of a time continuous model leads to the parameter $\Delta t$ in the equations above and offers therefore a direct way to deal with the gaps in Sgr~A*'s light curve for the conditional distribution. However, the transition matrix  $P = p_{ij}$ is calculated for a specific time difference $\tau$ and must also be modified when gaps are present. A straightforward way is to multiply the transition matrix with itself $N$ times for a gap that is $ \tau \cdot N$:
\begin{equation}
A = P^N, \textnormal{with } N = \max (1, \textnormal{round}(\Delta t / \tau)).
\end{equation}
For Sgr~A*'s data set, that has an average sampling of one measurement per 1.2 mins (not counting the big nightly / yearly gaps), we chose a final value of $\tau = 1$ min. We have explored much shorter values but found it to make no significant difference. For values of $\Delta t > 1000$ mins we set $\Delta t = 1000$ mins, since these gaps are safely greater than the coherence time scale of Sgr~A* \citep{meyer09,witzel12}. 

The stochastic model in equation~(\ref{OU}) aims to describe the intrinsic properties of the source under consideration. In a realistic setting, however, an additional noise component is present that reflects the measurement process. This measurement noise is typically white noise, i.e. it does not depend on the previous observation, and it is well described by a Gaussian and therefore fully specified by one parameter $\sigma_{\rm{meas}}$. Often, there exists an estimate of $\sigma_{\rm{meas}}$. For the case of Sgr~A* for example, nearby stars of similar magnitude visible in the same image offer a straightforward way to estimate the measurement noise, since these stars have constant flux intrinsically. In case an independent estimate of $\sigma_{\rm{meas}}$ is present, it can be advantageous to include it in the stochastic model of the source. In order to do this, the conditional distribution from the OU process has to be convolved with a Normal distribution $\mathcal{N} (0,\sigma_{\rm{meas}})$,
\begin{equation} \label{conv}
\tilde{f}(y_{t+\Delta t} | y_t;\sigma_{\rm{meas}}) = \int f(x|y_t) \cdot \frac{1}{\sqrt{2\pi \sigma_{\rm{meas}}^2}} \exp \left( \frac{-(y_{t+\Delta t} - x)^2}{2\sigma_{\rm{meas}}^2} \right) \; dx.
\end{equation}
For the important cases of a $\log$- or $\log\log$-OU process (eqs. \ref{logOU} and \ref{loglogOU}) this integral has to be calculated numerically in every term of the sum in the likelihood function eq.~(\ref{likeli}). Note that this dramatically increases the computing time. For us, the duration of computing the posterior with the numerical integration increased to the order of days from just a few hours without it. Please also note that this approach of incorporating the measurement noise assumes that it is constant. While photon noise leads to an increase of noise with flux, the data properties are well modeled with the assumption of constancy and may reflect the dominance of PSF measurement noise \citep[][Witzel et al., in prep.]{witzel12}.

\subsection{Is an additional state justified?} \label{just}
An important question is how to decide whether more than one state is needed at all, and if so, how many different states can be inferred from the data. Here lies an important distinction between astrophysical and economic analyses: while the latter are mainly interested in a precise model of a time series to make accurate forecasts, the former are insight driven. The necessity of different states could point to different physical mechanisms and elucidate the astrophysics of the source under consideration. Whether an additional state is necessary in the model of an astronomical time series should be assessed by several metrics. Bayesian methods like comparing the Bayesian evidences for different models belong to the standard methods well suited to compare models, but have the (dis-)advantage of forcing one to write down  specific priors for the parameters, which is often ambiguous. Note that \citet{hamilton05} warns that methods relying on likelihood ratio tests fail to satisfy necessary regularity conditions. 

A quite robust way of assessing the necessity of an additional state is to look at the global likelihood/posterior distribution in another way: too many assumed distinct states will lead to a highly multi-modal, very irregular looking posterior distribution. If only the maximum of the distribution is tried to be determined, the optimizer might often fail to converge at all in this case. Most importantly, the individual states should show persistence, if they are real. In any solution with a superfluous additional state the probability of remaining in that state $p_{22}$ will be $\ll 0.1$, while a significant additional state should show persistence with $p_{22} \gtrsim 0.8$. If the additional state has a very low probability of remaining in that state, two things might occur: (1) the probability of remaining in the first state is also very low, meaning that the states fluctuate from point to point. It seems extremely unnatural that state changes in the observed source occur exactly at the sampling rate of the measurements. (2) The probability of remaining in the first state is very high. In this case, the source will be in the additional, second state hardly at all. Only very few flux points will be assigned to that state and a natural explanation is that these are outliers for the assumed first state conditional distribution. This can easily happen if any assumption of the stochastic model is not quite accurate, e.g. if the flux density distribution is not quite log-normal or the measurement noise is not strictly constant.

\section{Results for Sgr~A* data}
In this section we will show the results of the regime-switching approach applied to Sgr~A*. We will use two data sets, both taken with adaptive optics in the near-IR K-band: all publicly available (up to 2010) VLT data as published in \citet{witzel12}, and all AO Keck data taken from Sgr~A*  (up to and including 2013; see Witzel et al., in prep.).  The photometry has been extracted in the same way in both data sets. We refer the reader to \citet[][and in prep.]{witzel12} for details. Key features of each data set are: 
\begin{itemize}
\item The VLT data were taken with NaCo in Ks-band (2.18 $\mu$m; 68 mas resolution), the Keck data with NIRC2 in K'-band (2.12 $\mu$m; 53 mas resolution),
\item the VLT data contain 10,639  quality-selected data points, taken between 6-13-2003 and 6-16-2010, and the Keck data contain 3,157 quality-selected points between 7-16-2004 and 7-19-2013, 
\item the average sampling of the covered time periods is 1.2 minutes (VLT) and 1.1 minutes (Keck), 
\item the integration time is 28 seconds for Keck and ranges between 30 and 40 seconds for VLT,
\item both data sets use consistent flux density calibration using 13 non-variable stars throughout all epochs, 
\item both data sets are corrected for extinction with $m_{ext}=2.46$ and for confusion levels  (epoch by epoch), 
\item the (Gaussian) measurement noise is determined to be 0.32 mJy (VLT) and 0.16 mJy (Keck),
\item typical background fluxes are 0.6 mJy (VLT) and 0.3 mJy (Keck), and
\item the data cover a de-reddened flux density range of 0--29 (VLT) / 23 (Keck) mJy which is  consistent with a power-law distribution of intrinsic fluxes in both cases.
 \end{itemize}
 
\begin{figure}
\includegraphics[scale=.75]{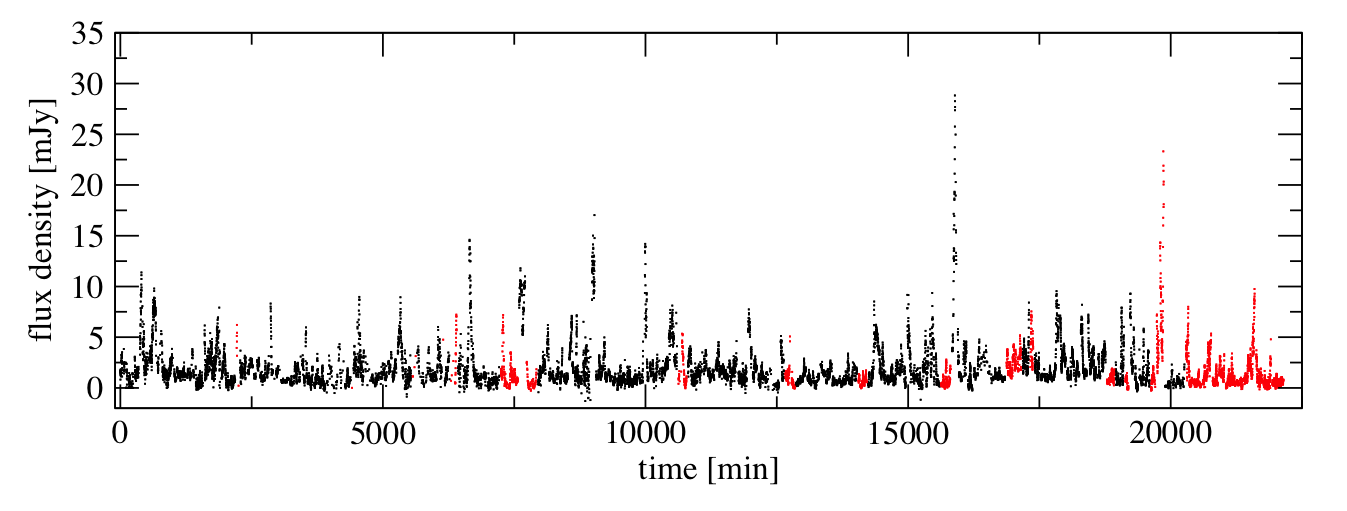}
\caption{This figure shows the longest, most comprehensive NIR light curve of Sgr A* that is available today ($13,800$ data points, here displayed without gaps longer than 30 min); see Witzel et al. (in prep.). This data has been taken in the K-band from 2003 to 2013 with both the VLT (black) and the Keck observatory (red) and shows a typical cadence of about 1 minute for the individual night. }
\label{lc}
\end{figure}

Figure~\ref{lc} shows the complete data set. Since the two data sets come from different telescopes and instruments, and show substantial differences in noise characteristics, we mainly analyze them separately.
 
In the following, we will first model SgrA* using a single state only. This will serve as our baseline model. We will then go on modeling the data with two states and see whether a substantial improvement has been achieved. 

\subsection{The role of measurement noise}

We will first present an analysis without accounting for the measurement (white) noise component. Since measurement noise is of course present in our data, this  might seem odd. However, we think that any first pass of a new data set should be done without the very time consuming convolution in equation (\ref{conv}), and we want to present our results in a way that the reader might and should approach his or her own data. In addition to avoiding the substantial computing time, there is another argument for not including the measurement noise at first is a meaningful approach: Since the approximately white measurement noise is very different from a damped random walk (in addition to a different mean and variance, it does not depend on the prior flux point and its mean reversion time scale is therefore infinity), any changes in the flux densities that are dominated by the measurement noise will be picked up as a second state in our algorithm. If a second state is convincingly found and that state is consistent with the white measurement noise, it is clear that no second state intrinsic to the source can be derived from the data. The measurement noise is a manifestation of the limit of what we can learn from given data and while it can be included in a model, it can not be removed. The treatment with the convolution  in equation (\ref{conv}) is still necessary to infer the accurate parameters of the model, but the simple question of how many variability states are present in a source can be answered without it.

\subsection{Baseline model}

Up to now, the timing behavior of Sgr~A* has always been modeled using only a single state in the literature. An advantage of this approach is that it allows for any choice of the overall flux distribution, e.g. a  power-law convolved with a Gaussian as in \citet{witzel12}. In our methodology, however, where we explicitly model conditional distributions, we do not have the freedom to choose a power-law behavior, since a dynamic model similar to eq.~(\ref{OU}) that results in a power-law distributions is not known.  We therefore have to show first that a $\log$- or $\log\log$-OU process can accurately model the overall, unconditional flux distribution as well. If this was not the case, any additional state would likely be preferred just because a single state does not appropriately reflect the distribution of fluxes. The key feature we are looking for is a different timing behavior, since this is expected from a significantly distinct state that reflects a distinct physical process in the accretion flow.

\begin{figure}
\includegraphics[scale=.75]{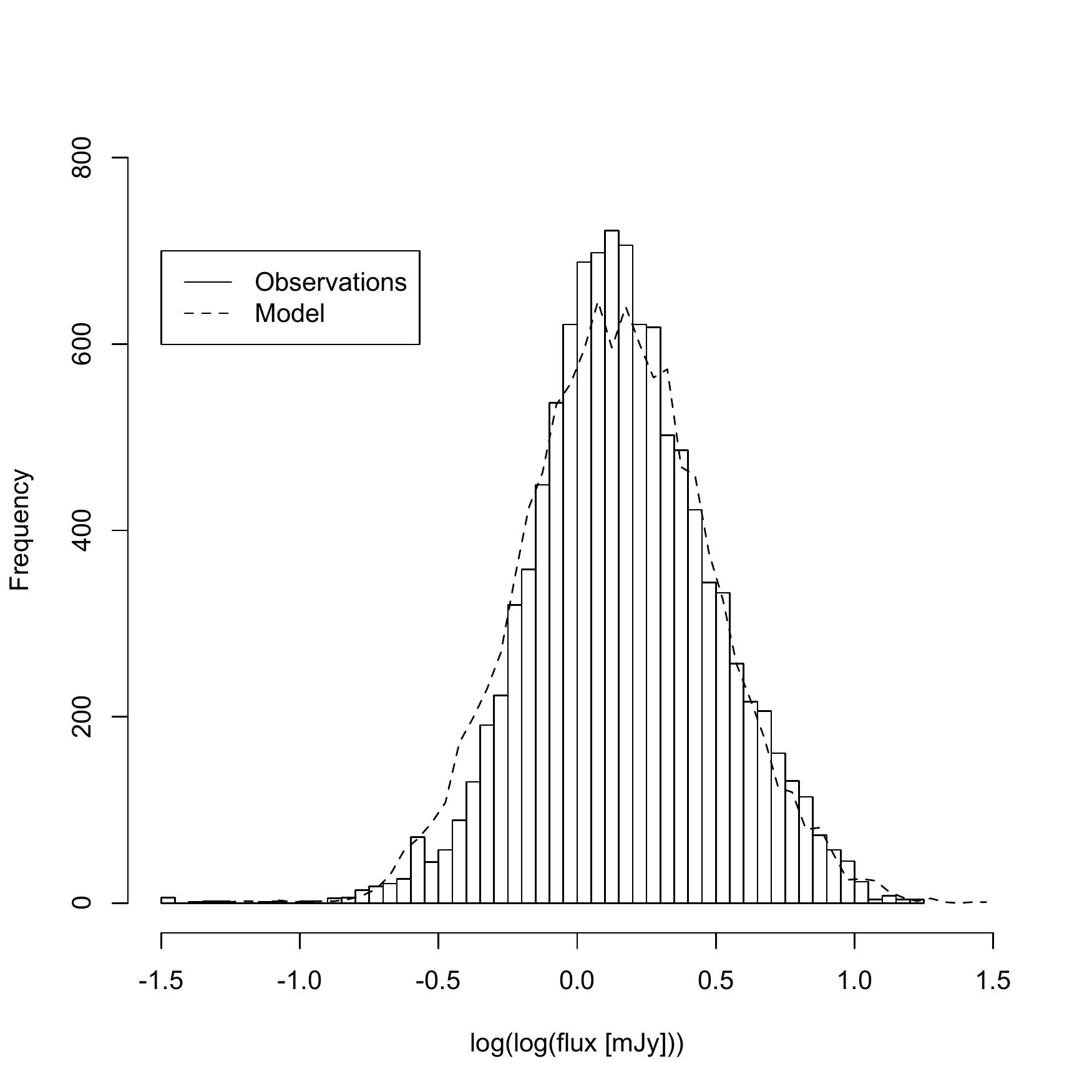}
\caption{The overall flux distribution of Sgr~A* from the VLT data set presented in \citet[][solid line]{witzel12} and our single state baseline model (dashed line). Concretely, the $\log\log$ of the flux after a constant value of 1.25 mJy has been added is shown. The observed distribution closely resembles a Gaussian, which shows that our assumptions about the baseline model are accurate.}
\label{hist}
\end{figure}

To first order, the total flux distribution of Sgr~A* is peaked and highly skewed. Exploring the distribution visually, it is noticeable that the  histogram of the $\log(\rm{flux})$ and $\log(\log(\rm{flux}))$ resembles a Gaussian, although still skewed in the latter case. This skew is minimized when a constant is added to the whole light curve, i.e. $y_t = y_t + c$ for all $t$, where $y_t$ denotes the flux density at time $t$. For the loglog of the flux this constant evaluates to $c = 1.25$ mJy for the VLT data and $c = 1.35$ mJy for the Keck data, and in the case of log(flux) it is 0 mJy for both data sets. Note that a constant of $c>1$ would be required in any way in order for the loglog to be defined, since the light curves are normalized such that $\min(y_t) = 0$. Table~\ref{tab1} shows the Bayesian evidences and the best fit parameter values for both models and data sets. The preferred model in both cases is the loglog one, although the difference to the log model is marginal for the VLT data. We adopt the loglog-OU process as the baseline when measurement noise is not included via eq.~(\ref{conv}), and the resulting model as well as the observed flux density histograms for the VLT data are shown in Fig.~\ref{hist}. The comparison of baseline model and data shows that while the agreement is very good, it is not perfect. However, since we can not use power-law distributions and do not explicitly model the measurement noise here, this is not surprising. 

We have used the Bayesian nested sampler {\it MultiNest} \citep{feroz08,feroz09} to explore the posterior distribution, and assumed uniform priors for the parameters $\mathcal{U}(0, 10)$.

\subsection{Modeling without the measurement noise component}

We can now turn to the question we set out to answer: does Sgr~A* have more than one state? While the choice of the conditional distribution for the baseline model was motivated to most closely match the observed overall flux distribution (for a single state the conditional flux distribution obviously equals the overall one), the two state case is more complex. Fundamentally, the selection of the form of the conditional distributions (e.g. OU, $log$-OU, ... ) is an assumption to make and can not be derived from the data themselves. In a practical approach, we have chosen many different combinations and looked at them individually.

The results of the modeling with different assumptions for the conditional distributions are summarized in table~\ref{tab1}. Strikingly, all two state scenarios are preferred over the single state model for both data sets as indicated by the Bayesian evidence. All posteriors are well behaved and not more than bi-modal (for a two-state model, the posterior will often be bi-modal), and the probabilities to remain in a given state is high, indicating persistence of these states. Taken together, this is strong evidence for the presence of more than one variability state in the observed flux from Sgr~A*. The overall preferred two-state model is one consisting of a $\log\log$-OU and a $\log$-OU state. This is true for both the VLT and Keck data sets.

\begin{figure}
\includegraphics[scale=.65]{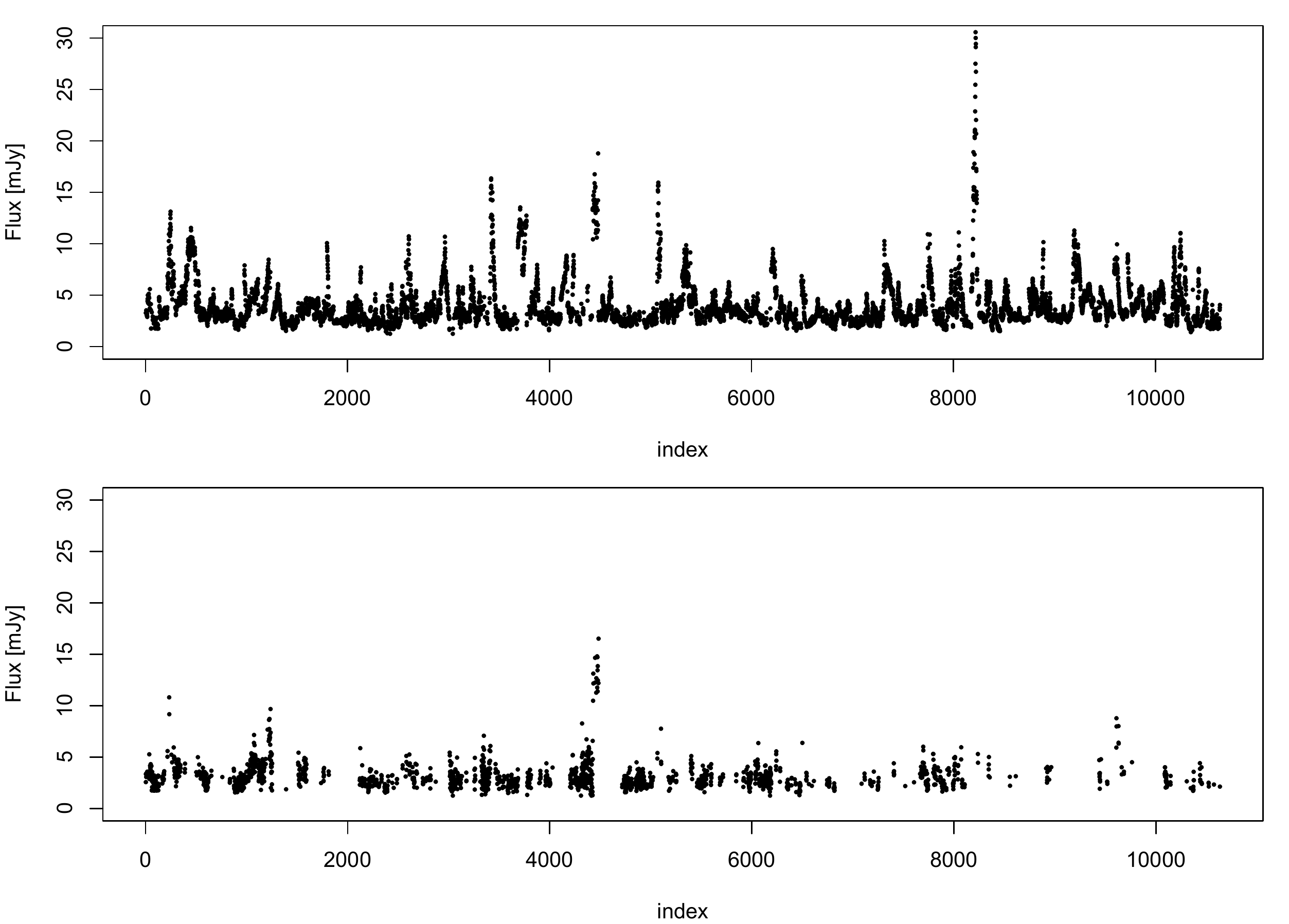}
\caption{Decomposition of the VLT data into the best-fit two state model (see Table~\ref{tab1}, row 4). {\it Upper panel:} All flux points of the time series that are with probability $> 0.5$ in the $\log$-OU state (83\% of all points).
 {\it Lower panel:} All flux points that are with probability $> 0.5$ in the $\log\log$-OU state (17\% of all points). This visualizes the differences in the two states: the $\log\log$-OU state is quickly reverting to a rather low mean, while the  $\log$-OU state is more slowly mean-reverting and the mean is higher. The $\log\log$-OU state represents the state where the changes in flux are dominated by measurement noise, and the $\log$-OU state represents the changes in flux that are source-dominated.}
\label{fig2}
\end{figure}

In Fig.~\ref{fig2} we show the observed VLT flux density time series in a decomposed way: in the upper panel are all points that are with probability $> 0.5$ in the $\log$-OU state, while the lower panel shows all flux points that are with probability $> 0.5$ in the $\log\log$-OU state. The differences are clearly visible. In fact, in all explored two-state models one state is always reverting to its mean fast and has a lower mean, while the other state is slower mean-reverting with a higher mean. A fast mean-reversion here means that this time scale is at the order of the sampling. We have also tested that the second state is not only preferred because of a better description of the overall flux distribution: forcing the two states to have the same form of the unconditional distribution by setting $\mu_1=\mu_2$ and  $\sigma_2 = k_2/k_1 \; \sigma_1$, we still see strong evidence for a second state (see last lines in Table~\ref{tab1}). It is the timing that drives the significance of a second state.

The fact that one of the two states is quickly mean reverting to a fairly low mean suggests the interpretation that this state is predominantly describing the parts of the light curve that are dominated by measurement noise, which is approximately white noise. Note that even if this is the case, the above analysis is still valuable since it offers a recipe to label each point-to-point change of the flux series as ``noise-dominated" or ``source-dominated". Consistent with this interpretation is the observation that the less noisy Keck data are only 6\% of the time ``noise-dominated", while the VLT data are  ``noise-dominated" 17\% of the time.

\begin{deluxetable}{lcc}
\tabletypesize{\scriptsize}
\tablewidth{0pt}
\tablecaption{Multi-state modeling of Sgr~A* (without measurement noise component)  \label{tab1}}
\tablehead{
        \colhead{Model} &
	\colhead{$\log$(Evidence)} &
	\colhead{Parameter values\tablenotemark{a}}
}
\startdata
\sidehead{VLT data}  
$\log\log$-OU & -9639 & $(k, \mu , \sigma) = (0.13, 0.16, 0.17)$ \\
$\log$-OU & -9642 & $(k, \mu , \sigma) = (0.10, 0.69, 0.30)$ \\
\hline
\noalign{\vskip .5ex}
$\log\log$-OU/OU & -5364 & $(k_1, \mu_1 , \sigma_1, p_{11}, k_2, \mu_2, \sigma_2, p_{21}) = (0.016, 0.24, 0.061, 0.96, 0.52, 2.90, 0.84, 0.11)$ \\
 \bf loglog-OU/log-OU & \bf -4813 &  $(k_1, \mu_1 , \sigma_1, p_{11}, k_2, \mu_2, \sigma_2, p_{21}) =(0.52, 0.016, 0.32, 0.91, 0.017, 1.29, 0.079, 0.02)$  \\
$\log\log$-OU/$\log\log$-OU & -5177 &  $(k_1, \mu_1 , \sigma_1, p_{11}, k_2, \mu_2, \sigma_2, p_{21}) =(0.019, 0.25, 0.059, 0.97, 0.42, 0.002, 0.28, 0.08)$ \\
$\log\log$-OU/$\log\log$-OU\tablenotemark{b} & -5334 &  $(k_1, \mu , \sigma, p_{11}, k_2, p_{21}) = (0.031, 0.16, 0.060, 0.97, 0.14, 0.08)$ \\
\sidehead{Keck data}
$\log\log$-OU & -733 & $(k, \mu , \sigma) = (0.04, 0.023, 0.12)$ \\
$\log$-OU & -939 & $(k, \mu , \sigma) = (0.04, 0.29, 0.26)$ \\
\hline
\noalign{\vskip .5ex}
$\log\log$-OU/OU & -158 & $(k_1, \mu_1 , \sigma_1, p_{11}, k_2, \mu_2, \sigma_2, p_{21}) = (0.02, 0.03, 0.08, 0.92, 0.016, 3.86, 0.33, 0.13)$ \\
\bf loglog-OU/log-OU & \bf -32 &  $(k_1, \mu_1 , \sigma_1, p_{11}, k_2, \mu_2, \sigma_2, p_{21}) =(0.17, 0.04, 0.29, 0.74, 0.015, 1.06, 0.073, 0.02)$ \\
$\log\log$-OU/$\log\log$-OU & -180 &  $(k_1, \mu_1 , \sigma_1, p_{11}, k_2, \mu_2, \sigma_2, p_{21}) =(0.017, 0.10, 0.067, 0.96, 0.10, 0.014, 0.21, 0.16)$ \\
$\log\log$-OU/$\log\log$-OU\tablenotemark{b} & -175 &  $(k_1, \mu , \sigma, p_{11}, k_2, p_{21}) = (0.021, 0.036, 0.067, 0.96, 0.067, 0.15)$ 
 \enddata
\tablenotetext{a}{The unit for the $k$ parameter is always $\rm{min}^{-1}$. The units for $\mu$ and $\sigma*\sqrt{t}$ depend on the model and are either mJy, $\log$(mJy), or $\log(\log$(mJy)). The prior for all parameters is a uniform distribution $\mathcal{U}(0, 10)$.}
\tablenotetext{b}{This model only allows for changes in the timing behavior. Here, we set $\mu_1 = \mu_2$ and $\sigma_2 = k_2/k_1 \; \sigma_1$. This tests whether the additional state is mainly driven by a different timing behavior (manifested in the mean reversion time $k$), or by the unconditional flux distribution (described by $\mu$ and $\sigma$).}
\end{deluxetable}

\subsection{Modeling with measurement noise}

Our hypothesis is that the quickly mean reverting state represents the measurement, white noise process. We will test this in two ways: since we have an estimate of $\sigma_{\rm{meas}}$ we can make use of eq.~(\ref{conv}). Furthermore, \citet{witzel12} developed a Monte Carlo tool to simulate Sgr~A* light curves (assuming a single state only),  and we can analyze these mock data separately. 

We summarize our results in Table~\ref{tab2} for the real data, and in Table~\ref{tab3} for the mock data. Regarding the real data, the evidence for a second state vanishes when the measurement noise is incorporated into the model. For both data sets the second state becomes highly non-persistent with a probability to remain in that state of $p_{22} \lesssim $1\%. This is in stark contrast to the case of modeling without the extra noise component where generally $p_{22} \gtrsim $90\%. For the VLT data, there is also no improvement in the Bayesian evidence. While there is some improvement in the evidence for the Keck data, the structure of the solution suggests that a few outliers drive this behavior (see also discussion in section~\ref{just}).

Regarding the mock data, table~\ref{tab3} shows that the mock data lead to a very similar solution as the real data, which is encouraging. Since the mock data algorithm was calibrated to the VLT data in terms of the sampling function and applied noise, the agreement between the mock data and the real VLT data is stronger than to the Keck data. When the measurement noise is not incorporated into the modeling, a second state is clearly preferred. Since the mock data follow a single-state process by construction, this shows that the white measurement noise leads to a discernible state in addition to the intrinsic red noise process of Sgr~A*.  And again, when the noise is explicitly convolved into the conditional distributions, the evidence for a second state vanishes. The solution puts the probability to remain in the second state to $p_{22} \lesssim  1\%$.  

In summary, the evidence for two distinct states in the flux from Sgr~A* disappears when the measurement noise is taken into account. We therefore see our hypothesis verified: the quickly mean-reverting, lower-mean state represents the state when the observed flux changes are dominated by instrumental noise, and the slowly mean-reverting, higher-mean state represents the state when the flux changes are dominated by Sgr~A* itself. 

It is interesting to note that the same result holds true when the analysis is done on a combined Keck and VLT data set (see Fig.~\ref{lc}). While two states are again being picked up when the measurement noise is unaccounted for, one state tends to describe white noise and the other Sgr~A* intrinsically. With the measurement noise convolved into the conditional distributions, the evidence for a second state vanishes.

\begin{deluxetable}{lcc}
\tabletypesize{\scriptsize}
\tablewidth{0pt}
\tablecaption{Multi-state modeling of Sgr~A* with measurement noise component\tablenotemark{a}  \label{tab2}}
\tablehead{
        \colhead{Model} &
	\colhead{$\log$(Evidence)} &
	\colhead{Parameter values\tablenotemark{b}}
}
\startdata
\sidehead{VLT data}  
$\log\log$-OU & -7899 & $(k, \mu , \sigma) = (0.06, 0.14, 0.10)$ \\
$\log$-OU & -7436 & $(k, \mu , \sigma) = (0.04, 0.75, 0.17)$ \\
\hline
\noalign{\vskip .5ex}
 loglog-OU/log-OU & -7420 &  $(k_1, \mu_1 , \sigma_1, p_{11}, k_2, \mu_2, \sigma_2, p_{21}) =(0.07, 0.17, 0.10, 0.99, 0.72, 1.39, 0.56, 0.98)$  \\
  &  & \it 2nd state irrelevant and non-persistent ($p_{22} = $2\%)\\
\sidehead{Keck data}
$\log\log$-OU & -432 & $(k, \mu , \sigma) = (0.02, 0.04, 0.07)$ \\
$\log$-OU & -417 & $(k, \mu , \sigma) = (0.02, 0.37, 0.16)$ \\
\hline
\noalign{\vskip .5ex}
loglog-OU/log-OU &  -194 &  $(k_1, \mu_1 , \sigma_1, p_{11}, k_2, \mu_2, \sigma_2, p_{21}) =(0.01, 0.04, 0.05, 0.97, 0.20, 1.33, 0.28, 0.99)$  \\
  &  & \it 2nd state irrelevant and non-persistent ($p_{22} = $1\%)\\
 \enddata
\tablenotetext{a}{The value for $\sigma_{\rm{meas}}$ is $\sigma_{\rm{meas}} = 0.32$ mJy for the VLT data \citep{witzel12}, and $\sigma_{\rm{meas}} = 0.16$ mJy for the Keck data (Witzel et al., in prep.).}
\tablenotetext{b}{The unit for the $k$ parameter is always $\rm{min}^{-1}$. The units for $\mu$ and $\sigma*\sqrt{t}$ depend on the model and are either $\log$(mJy) or $\log(\log$(mJy)). The prior for the parameters is a uniform distribution $\mathcal{U}(0, 1)$, only for $\mu_2$ it is $\mathcal{U}(0, 3)$.}
\end{deluxetable}

\begin{deluxetable}{lcc}
\tabletypesize{\scriptsize}
\tablewidth{0pt}
\tablecaption{Multi-state modeling of mock data\tablenotemark{a}  \label{tab3}}
\tablehead{
        \colhead{Model} &
	\colhead{$\log$(Evidence)} &
	\colhead{Parameter values\tablenotemark{b}}
}
\startdata
\sidehead{Modeling without measurement noise component}  
$\log\log$-OU & -9157 & $(k, \mu , \sigma) = (0.10, 0.15, 0.15)$ \\
$\log$-OU & -10477 & $(k, \mu , \sigma) = (0.11, 0.87, 0.29)$ \\
\hline
\noalign{\vskip .5ex}
 loglog-OU/log-OU &  -7975 &  $(k_1, \mu_1 , \sigma_1, p_{11}, k_2, \mu_2, \sigma_2, p_{21}) =(1.25, 0.001, 0.28, 0.97, 0.04, 1.48, 0.11, 0.01)$  \\
\sidehead{Modeling with measurement noise component\tablenotemark{c}}
$\log\log$-OU & -8486 & $(k, \mu , \sigma) = (0.04, 0.25, 0.07)$ \\
$\log$-OU & -8286 & $(k, \mu , \sigma) = (0.04, 0.87, 0.14)$ \\
\hline
\noalign{\vskip .5ex}
loglog-OU/log-OU &  -8248 &  $(k_1, \mu_1 , \sigma_1, p_{11}, k_2, \mu_2, \sigma_2, p_{21}) =(0.03, 0.04, 0.06, 0.50, 0.07, 1.75, 0.09, 0.999)$  \\
  &  & \it 2nd state non-persistent ($p_{22} = $0.1\%)\\
 \enddata
\tablenotetext{a}{The model for the mock data is taken from \citet{witzel12} where it has been calibrated with the VLT data and uses a single state only.}
\tablenotetext{b}{The unit for the $k$ parameter is always $\rm{min}^{-1}$. The units for $\mu$ and $\sigma*\sqrt{t}$ depend on the model and are either $\log$(mJy) or $\log(\log$(mJy)). When no measurement noise is modeled, the prior for all parameters is a uniform distribution $\mathcal{U}(0, 10)$. When measurement noise is modeled, the prior for the parameters is a uniform distribution $\mathcal{U}(0, 1)$, only for $\mu_2$ it is $\mathcal{U}(0, 3)$.}
\tablenotetext{c}{The value for $\sigma_{\rm{meas}}$ is $\sigma_{\rm{meas}} = 0.32$ mJy.}
\end{deluxetable}

\subsection{The parameters of the single-state (best-fit) model} \label{35}

We have shown that Sgr~A* is sufficiently described by a single-state process once the measurement noise is accounted for in the modeling. The accurate model for Sgr~A*'s light curve is therefore the log-OU process convolved with Gaussian noise (see table~\ref{tab2}). While it is tempting to interpret these parameters and compare them to different modeling approaches of the past, one has to be cautious when the single-state parameter inference is done as the limiting case of a multi-state model, as we have done here. 

The reason for caution is the following: Our approach models the conditional distribution from one measurement to the next, $f(y_{t+\Delta t} | y_t)$, which is necessary in the context of a multi-state Markov model. If only a single state is assumed, this means, however, that only the typical sampling horizon $ \Delta t$ is used to estimate the mean-reversion time scale $k$. If the OU process is completely accurate and describes the true nature of the source, this is unproblematic. If, however, there are deviations from the OU process in the data, a different sampling horizon would lead to a different estimate of $k$. Ideally, a global approach using all time lags in the data would be used to estimate $k$ for a single-state process. Since we focus here on multi-state modeling, this is beyond the scope of this work. We would like to note, however, that the analysis of our mock data lead to consistent results with the analysis of the real data, which means we explicitly confirm the results of \citet{witzel12}.

\section{Discussion \& Conclusion}

Using the most extensive light curve of Sgr~A* to date with combined data from the VLT and Keck Observatory (Witzel et al., in prep.), we arrive at two main results: (1) the observed flux from Sgr~A* shows two distinct states, a noise-dominated and a source-dominated one, and (2) the intrinsic variability of Sgr~A* is sufficiently described by a single state stochastic process. Both findings have interesting implications, and we will discuss them in turn.

The presence of a noise-dominated and a source-dominated state in Sgr~A* is a manifestation of the colloquial language used in the Galactic center community, where it is often talked about a quiescent and a flaring state of Sgr~A* in the NIR. We have shown here that this intuitive distinction is right when describing observed light curves, however Sgr~A* is intrinsically not in quiescent and flaring states but rather in one single state only. The language is therefore somewhat imprecise and the word ``flare'' should not be used to describe Sgr~A* in the NIR. This also means that the lack of statistical evidence for a QPO of $\sim 17$ minutes can not be explained by distinct states of which  only one is accompanied by a QPO, a possibility raised by \citet{genzel10}. The fact that the intrinsic behavior of Sgr~A* shows no evidence for a second state explores and falsifies the idea first suggested by \citet{dodds11}. Physically, this means that one stochastic process without the addition of time resolved, discrete events -- like the disruption of asteroids -- fully explains our data set. 

It will be interesting to see whether a change of Sgr~A*'s intrinsic variability state changes when the gaseous, red, emission-line object G2 has passed the black hole in early 2014 \citep{gillessen12,gillessen13a,gillessen13b,phifer13,meyer13}. Since there is gas associated with this object, it could potentially offer a unique probe to observe the response of the accretion flow to a sudden increase of mass. However, the amount of gas is uncertain and it is unclear if there will be any visible effect to the flux emission from Sgr~A*. Two scenarios are likely; either something obvious will happen to Sgr~A*'s flux distribution like a shift towards a much higher mean, or -- if there is any change at all -- it can be very subtle requiring a detailed statistical method like the one we have developed here. This statistical methodology, in addition to the best possible data baseline by merging the VLT and Keck data, might prove crucial in understanding G2's impact on Sgr~A*.

Compared to simpler methods, our Hidden Markov Model approach presented here has the distinct advantage that it not only answers the question whether two states are needed, but it also solves for the times when the state changes happen. For G2 this is important, because although the time and distance of closest approach is fairly well known for G2's orbit, it is unclear when the gas has come down from $\sim 2400 R_S$ all the way to a few $R_S$ where the near-infrared flux gets emitted. Given enough observations, solving for the time of state change (if there will be one) can directly test accretion flow dynamics around Sgr~A*.  

Another benefit of our approach that assigns a probability to each flux change to be noise- or source-dominated is that it offers a recipe to get an astrometric position of Sgr~A* in the infrared. The astrometry of stars orbiting around the black hole aims to detect deviations from a purely Keplerian orbit as the next milestone \citep[e.g][]{ghez08,gillessen09,meyer12}. The astrometry is currently limited by the construction of an absolute reference frame that is used to transform all star positions into a common frame. This frame of reference is defined by seven maser sources visible in the radio as well as infrared \citep{yelda10}. Since Sgr~A* is also visible at both of these wavelengths and sits at a position that is most important to anchor the reference frame, it could be of tremendous help in overcoming the current limitations. However, reliable astrometry of Sgr~A* has been elusive. Its position changes as a function of brightness, probably because unresolved stellar sources bias its position, and the resulting astrometric shift is dependent on the relative contribution of Sgr~A*'s intrinsic flux. Our two-state modeling approach effectively filters out the flux point variations in the time series that are dominated by Sgr~A* itself. Getting a position based on the images only where the apparent flux from Sgr~A* is source-dominated should improve the astrometry and lead to a consistent determination of its position. 

As a final remark we would like to note that the multi-state methodology present here is directly applicable to other sources such as AGN as well. The only key assumption is the validity of the OU process (also called a damped random walk). This also means that the method is not wavelength specific. The analysis of optical AGN light curves, e.g., should be possible in exactly the same way as is presented here.  


\acknowledgments
We would like to thank Eric Becklin, Tuan Do, Matt Malkan, and Mark Morris for very valuable comments and discussions. Support for this work was provided by UCLA's OVCR-COR Transdisciplinary Seed Grants, NSF grant AST-0909218, and the Levine-Leichtman Family Foundation. Data presented herein were taken at the Keck Observatory, which is operated as a scientific partnership among Caltech, UC, and NASA; the Observatory was made possible by the generous financial support of the Keck Foundation. This paper is partially based on observations conducted with the European Southern Observatory telescopes obtained from the ESO/STECF Science Archive Facility.

{\it Facilities:} \facility{Keck:II}, \facility{VLT:Yepun}

\clearpage

\begin{thebibliography}{}

\bibitem[Aschenbach et al.(2004)]{aschenbach04} Aschenbach, B., Grosso, N., Porquet, D., \& Predehl, P.\ 2004, \aap, 417, 71

\bibitem[Baganoff et al.(2001)]{baganoff01} Baganoff, F. K., Bautz, M. W., Brandt, W. N., et al. 2001, Nature, 413, 45

\bibitem[Balick \& Brown(1974)]{balick74} Balick, B., \& Brown, R.~L.\ 1974, \apj, 194, 265 

\bibitem[Blandford \& Begelman(1999)]{blandford99} Blandford, R.~D., \& Begelman, M.~C.\ 1999, \mnras, 303, L1

\bibitem[Broderick \& Loeb(2005)]{broderick05} Broderick, A.~E., \& Loeb, A.\ 2005, \mnras, 363, 353  

\bibitem[Dexter et al.(2013)]{dexter13} Dexter, J.,  Kelly, B. C., Bower, G. C., et al. 2013, subm. to MNRAS, eprint arXiv:1308.5968

\bibitem[Do et al.(2009)]{do09} Do, T., Ghez, A. M., Morris, M. R., et al. 2009, ApJ, 691, 1021


\bibitem[Dodds-Eden et al.(2011)]{dodds11} Dodds-Eden, K., et al. 2011, \apj, 728, 37

\bibitem[Eckart et al.(2004)]{eckart04} Eckart, A., Baganoff, F.~K., Morris, M., et al.\ 2004, \aap, 427, 1 


\bibitem[Falcke \& Markoff(2013)]{falcke13} Falcke, H., Markoff, S. B. 2013, CQGra, 30, 4003

\bibitem[Feroz, Hobson \& Bridges(2009)]{feroz09} Feroz, F., Hobson, M. P., Bridges, M. 2009, MNRAS, 398, 1601

\bibitem[Feroz \& Hobson(2008)]{feroz08} Feroz, F., Hobson, M. P. 2008, MNRAS, 384, 449

\bibitem[Genzel et al.(2003)]{genzel03}
  Genzel, R., Sch\"odel, R., Ott, T., et al. 2003, Nature, 425, 934
  
\bibitem[Genzel, Eisenhauer \& Gillessen(2010)]{genzel10} Genzel, R., Eisenhauer, F., Gillessen, S. 2010, Reviews of Modern Physics, 82, 3121 

\bibitem[Ghez et al.(2004)]{ghez04} Ghez, A. M., Wright, S. A., Matthews, K., et al., 2004, \apj, 601, L159

\bibitem[Ghez et al.(2008)]{ghez08} Ghez, A. M., Salim, S., Weinberg, N. N., et al. 2008, \apj, 689, 1044

\bibitem[Gillessen et al.(2009)]{gillessen09}Gillessen, S.,  Eisenhauer, F., Trippe, S. et al. 2009, \apj, 692, 1075

\bibitem[Gillessen et al.(2012)]{gillessen12} Gillessen, S., Genzel, R., Fritz, T. K., et al. 2012a, Nature, 481, 7379, 51

\bibitem[Gillessen et al.(2013a)]{gillessen13a} Gillessen, S., Genzel, R., Fritz, T. K., et al. 2013, \apj, 763, 78

\bibitem[Gillessen et al.(2013b)]{gillessen13b} Gillessen, S., Genzel, R., Fritz, T. K., et al. 2013, submitted to \apj, eprint arXiv:1306.1374

\bibitem[Hamilton(1994)]{hamilton} Hamilton, J. D., 1994, ÒTime Series AnalysisÓ, Princeton University Press, Princeton, New Jersey

\bibitem[Hamilton(2005)]{hamilton05} Hamilton, J. D., 2005, ÒRegime-Switching ModelsÓ, Palgrave Dictionary of Economics

\bibitem[Johannsen \& Psaltis(2011)]{johannsen11} Johannsen, T., \& Psaltis, D.\ 2011, \apj, 726, 11 

\bibitem[Kelly et al.(2009)]{kelly09} Kelly, B. C., Bechtold, J., Siemiginowska, A. 2009, 698, 895


\bibitem[MacLeod et al.(2010)]{macleod10} MacLeod, C. L., Ivezic, Z., Kochanek, C. S., et al. 2010, 721, 1014


\bibitem[Meyer et al.(2006)]{meyer06b} Meyer L., Eckart A., Sch\"odel R., Duschl, W. J., Muzic, K., Dovciak, M., Karas, V., 2006b, A\&A, 460, 15 


\bibitem[Meyer et al.(2008)]{meyer08} Meyer, L., Do, T., Ghez, A., et al. 2008, \apj, 688, L17

\bibitem[Meyer et al.(2009)]{meyer09} Meyer, L., Do, T., Ghez, A., et al. 2009, \apj, 694, L87

\bibitem[Meyer et al.(2012)]{meyer12} Meyer, L., Ghez, A. M., Sch\"odel, R., et al. 2012, Science, 338, 6103, 84

\bibitem[Meyer et al.(2013)]{meyer13} Meyer, L., Ghez, A. M, Witzel, G., et al. 2013, Proceedings of IAU Symposium \#303, eprint arXiv:1312.1715

\bibitem[Morris, Meyer \& Ghez(2012)]{morris12} Morris, M. R., Meyer, L., Ghez, A. 2012, RAA, 12, 995

\bibitem[Narayan et al.(1995)]{narayan95} Narayan, R., Yi, I., \& Mahadevan, R.\ 1995, \nat, 374, 623 

\bibitem[Neilsen et al.(2013)]{neilsen13} Neilsen, J., Nowak, M.~A., Gammie, C., et al.\ 2013, \apj, 774, 42

\bibitem[Phifer et al.(2013)]{phifer13} Phifer, K., Do, T., Meyer, L., et al. 2013, \apjl, in press, eprint arXiv:1304.5280

\bibitem[Rabiner(1989)]{rabiner89} Rabiner, L. 1989, Proceedings of the IEEE, 77, 2, 257

\bibitem[Witzel et al.(2012)]{witzel12} Witzel, G., Eckart, A., Bremer, M., et al. 2012, ApJS, 203, 18

\bibitem[Yelda et al.(2010)]{yelda10} Yelda, S., Lu, J. R., Ghez, A., et al. 2010, \apj, 725, 331

\bibitem[Yuan et al.(2003)]{yuan1} Yuan, F., Quataert, E., Narayan, R. 2003, ApJ, 598, 301

\bibitem[Yuan et al.(2004)]{yuan2} Yuan, F., Quataert, E., Narayan, R. 2004, ApJ, 606, 894

\bibitem[Zu et al.(2013)]{zu13} Zu, Y., Kochanek, C. S., Kozlowski, S., Udalski, A. 2013, ApJ, 765, 106


\end{thebibliography}
\end{document}